# ARTIFICIAL INTELLIGENCE ETHICS: AN INCLUSIVE GLOBAL DISCOURSE?

Cathy Roche, Science Foundation Ireland, CRT-AI, Trinity College Dublin, croche7@tcd.ie

Dave Lewis, ADAPT Centre, Trinity College Dublin, dave.lewis@adaptcentre.ie

P. J. Wall, ADAPT Centre, Trinity College Dublin, pj.wall@tcd.ie

**Abstract:** It is widely accepted that technology is ubiquitous across the planet and has the potential to solve many of the problems existing in the Global South. Moreover, the rapid advancement of artificial intelligence (AI) brings with it the potential to address many of the challenges outlined in the Sustainable Development Goals (SDGs) in ways which were never before possible. However, there are many questions about how such advanced technologies should be managed and governed, and whether or not the emerging ethical frameworks and standards for AI are dominated by the Global North. This research examines the growing body of documentation on AI ethics to examine whether or not there is equality of participation in the ongoing global discourse. Specifically, it seeks to discover if both countries in the Global South and women are underrepresented in this discourse. Findings indicate a dearth of references to both of these themes in the AI ethics documents, suggesting that the associated ethical implications and risks are being neglected. Without adequate input from both countries in the Global South and from women, such ethical frameworks and standards may be discriminatory with the potential to reinforce marginalisation.

## 1. INTRODUCTION

Increasing advancements in AI have changed how we interact, live and work. The unprecedented transformative potential of these technologies to address the many and varied challenges outlined in the SDGs, including the eradication of poverty, zero hunger and good health, cannot be ignored. Accompanying the evolution of the technology itself, there has been increasing discourse on the topic of AI ethics, with many principles, guidelines, frameworks, declarations, strategies, charters, policies and position papers being issued by a variety of agencies, non-governmental organisations (NGOs), and governments. These are important discussions, as both the AI technology itself and the ethical frameworks and standards emerging around it can reproduce and reinforce a variety of biases if not designed, developed and deployed on the basis of inclusive participation. If the voices of all affected communities are not included, then such ethical frameworks and standards can remain discriminatory and reinforce marginalisation (Molnar, 2020). As the body of AI ethics and governance documentation continues to expand, it is appropriate to examine and analyse this emerging literature from the perspective of inclusivity. In particular, it is important to examine AI ethical frameworks and guidelines through the lenses of gender and the Global South as women in such contexts are often the most impacted by technological developments while also being marginalised in their design, development and operation.

The existing body of documents on AI ethics has to date focussed on discerning concordance on ethical principles or differences in approach by organisational sectors. This research therefore seeks to address a gap, by examining the existing body of work for reference to gender and sustainability (serving as a proxy for inclusion of the Global South) in order to assess whether there is underrepresentation, both in terms of gender and geographic areas such as Africa, Central and South America, Asia and Oceania. Such underrepresentation is likely to be indicative of unequal participation in the debate around the ethics of AI, exposing an international discourse marked by power imbalance. Based on this, the following research question is posed: is there equality of participation in the global discourse on AI ethics, or are women and countries in the Global South underrepresented in this discourse?

Positioning this research broadly within the ICT4D field, it makes a specific contribution to the growing sub-fields of AI for global development (AI4D), gender in ICT4D, and the body of work concerned with the ethical implications of AI and other advanced technologies for global development (AIethics4D). This is an important contribution for many reasons: most importantly, if the ethical agenda around AI is being set by countries in the Global North, there exists the potential to compound inequalities and further embed colonial ideologies in the Global South.

The paper proceeds as follows: firstly, the bodies of literature concerning AI ethics and standards are considered before examining the body of work on AI and social structures. Section 3 then presents the research methodology adopted, with research findings to date presented in Section 4. The paper concludes with Section 5 where brief conclusions are presented.

## 2. LITERATURE REVIEW

As mentioned, there is a rapidly expanding body of work on the topic of AI ethics, principles and guidelines. This section commences with a brief examination of this work, before moving on to present the literature on social and power structures in AI.

### 2.1. AI Standards, Principles and Guidelines

A variety of studies have identified emerging commonalities and levels of consensus across the body of work concerned with AI principles and guidelines. A key study delineates a global convergence around key ethical principles such as justice and fairness, transparency, non-maleficence, privacy and responsibility (Jobin et al., 2019). At the same time, this study also highlights significant divergence around interpretation of these principles and their proposed implementation. In addition, other authors identify differences among the documentation associated with the provenance of the literature (Schiff et al., 2021). In their analysis of ethical topics across documents issued by private and public organisations as well as NGOs, they found greater ethical breadth and more engagement with law and regulation in the documents from NGOs and public sector issuers when compared to those from private sources.

### 2.2. Social and Power Structures

Regardless of origination in the public or private sector, any proposed ethical framework or governance model is reflective of power structures within the society in which they are developed. As technology is "*ultimately influenced by the people who build it and the data that feeds it*" (Chowdhury & Mulani, 2018), it is therefore reflective of the cultural and social biases of its context. By extension, ethical frames of reference and concerns are a product of their context and are therefore also subject to the prevailing culture and the risk of ethnocentrism.

In many ways, AI has augmented existing inequalities inherent in societal structures that are sexist and patriarchal but also racist and colonial. Concerned about a perceived structural domination by the United States (US) in the Global South, exercised through control of the digital ecosystem, Kwet (2019) describes an "*insidious new phenomenon, digital colonialism*", which is shaping the digital destiny of many African countries. US dominance of network connectivity, hardware and software in turn grants great economic and social power to large technology corporations, such as Microsoft, Apple and Google. Looking more closely at AI, Birhane posits the concept of algorithmic colonialism, wherein domination politically, economically and ideologically is achieved through approaches such as "*'technological innovation', 'state-of-the-art algorithms', and 'AI solutions' to social problems*" (Birhane, 2020). In the context of this perceived technological imperialism, it is of immense importance that those setting the ethical agenda around AI are cognisant of the potential to compound inequalities and further embed ideologies originating in the Global North. Indeed, any discussion of inequality and power, as relating to AI, cannot be ahistorical and will be incomplete if they do not take cognisance of "*colonial continuities*" (Mohamed et al., 2020).

As the impact of AI is not felt equally, such technologies embody the risk of further strengthening global digital inequality, especially amongst marginalised populations. Such groups encompass ethnic and racial groups, those with disabilities, young and LGBTQ people, poor rural and urban communities and especially women and those at the intersection of such identities. The prevalence of gender bias replicated in and by AI systems (Bolukbasi et al., 2016; Buolamwini & Gebru, 2018; Dastin, 2018) affects women globally but has the potential to be more damaging to women in the Global South. As highlighted in a recent report "*these gender biases risk further stigmatizing and marginalizing women on a global scale*" and that due to the ubiquity of AI in our lives, such biases put women at risk of being left behind "*in all realms of economic, political and social life*" (UNESCO, 2020). A complex interplay of issues exists at the intersection of AI ethics, the Global South and women. Therefore, in tandem with a concern for representation from the Global South more generally, should be one for the inclusion of women from the Global South in the global discourse around the ethics of AI.

## 3. RESEARCH METHODOLOGY

### 3.1. Document Collection

This study uses the same corpus as collected by Jobin et al. (2019) in their scoping review of existing non-legal norms or soft-law documentation. Comprising 84 sources or parts thereof, this collection includes policy documents, such as guidelines, institutional reports and principles but excludes legal and academic sources. Within the synthesised listing are sources written in English, French, German, Greek and Italian which explicitly reference AI in the title or description. All documents are considered to express a normative ethical stance defined as a "*moral preference for a defined course of action*" (Jobin et al., 2019). Further, the collection contains documents by issuing entities from both public and private sectors.

As there is no single database for AI ethics frameworks, principles or guidelines, the final synthesis of 84 documents garnered by Jobin et al. (2019) can be considered to constitute a corpus of ethical guidelines. This approach of using an externally defined collection also avoids any unconscious selection bias on the part of the authors of this study. In some instances, small deviations from this body of papers were necessary, where the original sources were no longer available at the given location or where a document has been superseded by a newer version. Where the document was unavailable, an equivalent source from the same issuer was selected and in the case of the new versions, the most recent was substituted thereby maintaining the integrity of the collection in terms of both content and issuing body. In all instances the volume remained the same (N=84). Documents were collected between January and February 2021, and Appendix A lists all documents/sites, issuing body and country of issuer.

To enable deeper sectoral analysis, the corpus was classified according to type of issuing entity. Categorisation as either 'NGO', 'Private' or 'Public' is an adaptation of the classification used in the Schiff et al. review of 112 AI global ethics documents (Schiff et al., 2021). Their findings revealed distinct differences in the handling of AI ethics by organisation type, with NGO and public documents both more participatory in origin and engaged with the law and private sources more focussed on ethical issues relating to customers or clients. Of particular interest to this study is the conclusion from that paper that NGO documents cover "*a range of ethical issues that are given less attention by other sectors*". It was therefore deemed of relevance to include such categorisation in this study to investigate if there were any sectoral dimensions to considerations of gender, feminism, the SDGs and the Global South in ethical guidance around AI.

| Organisation Category | Number of documents |
|---|---|
| NGO | 35 |
| Private | 18 |
| Public | 31 |

**Table 1 : Categorisation by Organisation Type**

As the Schiff et al. (2020) and the Jobin et al. (2019) papers analyse different corpora, the classification process was conducted manually for the 84 documents in this study. In terms of sectoral categorisation, 'Public' here includes sources issued by government entities and intergovernmental bodies, 'Private' contains for-profit companies and corporations while 'NGO' incorporates advocacy agencies, research groups, professional associations and academic collaborations. Based on the categorisation of the issuing entity or entities, one of the three labels in this taxonomy was assigned to each source and appended to the list in Appendix A. Table 1 shows the totals following classification of documents by sector/ issuer type.

### 3.2. Content Analysis and Coding

Content analysis of the documents was conducted in three main phases: the first involved coding for 'Sustainability' and related search terms before categorical (sectoral) analysis on the theme; the second coding cycle tackled analysis for 'Gender' and associated search terms and mapping by sector for this theme; a final phase incorporated comparative analysis of the two themes across documents and by issuer type. Consistency checks were performed throughout the analytical process by manual assessment of accuracy and reliability. This entailed researchers checking a random selection of the documents to assess if the software tool (see details below) was returning search terms correctly, finding all occurrences of given terms and not including terms outside the defined scope. As a result, coding underwent a process of refinement and was subsequently broadened or narrowed as required to ensure as many relevant terms as possible and their occurrences were captured. While not exhaustive, the final suite of terms used is broad and complete enough to achieve a thorough and robust examination for both themes.

Commencing with 8 simple terms that would evaluate if the documents took into consideration the concept of sustainability or a focus on the Global South, coding was amended through an iterative process. Adjustments to the coding were made to take account of a breadth of terms that could capture such concerns, such as 'developing economies' and 'emerging economies' as synonyms for developing countries. A list of all included codes is presented in Table 2. The codes are not case sensitive and take account of both American and English spellings. Also some applications of the terms are excluded due to being unrelated to the theme. For example, 'global justice', 'global gap' and 'global poverty' are counted under the "Global South" search term but 'globalisation' is excluded. Similarly, occurrences of 'third countries' are discounted as unrelated to the theme and having a very specific meaning in a European context. While the majority of the documents in the corpus were in English (either as language of origination or translated), one source in Italian and one in French were analysed in their original language, when translations could not be found.

Following finalisation of the coding for the Sustainability theme, data analysis was performed using R in RStudio version 1.4.1106 for Mac. This design decision is based on the potential for replicability of the research so that it can be repeated with additional corpora and at regular intervals, thereby generating a gender or inclusion observatory on AI. All documents were first converted to PDF format and web sources were saved down in MS Word before PDF conversion. Use of the R package 'pdfsearch' was considered the most appropriate tool as this package includes functions for keyword search of PDF files, but also provides a wrapper ('keyword_directory') that includes a function to loop over all PDF files within a single directory. Specifying multiple keywords for the search was achieved by creating a character vector. Other operations included ignoring case, removing hyphens and returning surrounding lines in addition to the matching line. As noted, outputs and results were checked manually for reliability and accuracy. Due to the complexity of incorporating multiple languages into the automated process, and given the majority of the documents (82 of 84) could be handled this way, English was set as the corpus language. Manual analysis of the French and Italian documents was performed.

The last step in this first phase of analysis takes the output of the keywords search for the Sustainability theme and compares results across categories, based on the 'NGO'/ 'Private'/ 'Public' taxonomy outlined earlier. Table 3 shows the classification of documents by sector/ issuer type. Categorical analysis was undertaken to identify any potential divergence across the documentation that might be attributable to the type of issuing organisation.

The second phase of content analysis comprised coding for the Gender theme before conducting the categorical analysis of search results for this theme. Again, starting with 8 simple terms that should assess the presence of the concepts of gender equality and feminism, after a small number of initial iterations, the coding was modified to include 'Feminisation'. This concept, while related to the other search terms, is distinct in the context of the feminisation of the workplace or the feminisation of personal assistant devices or robots. For the resultant 9 search terms, a list of included codes is given in Table 2. As described earlier, the codes again include English and American spellings and are not case sensitive but certain uses of the terms are deliberately excluded due to potential skewing of results in an over-represented way. For example, while 'sexual harassment', 'sexual violence' and 'sexualised' are included, 'sexuality' is not as this appears in the documentation solely in relation to sexual orientation rather than in terms of gender or gender identity. Similarly, occurrences of 'gendered' are counted but those of 'engendered' are excluded as this does not relate to the theme. The Italian and French documents were again evaluated manually while the majority of the corpus was analysed using R. Finally, for this phase, results from content analysis of this theme were compared on a sectoral basis.

Subsequently, a synthesis analysis of the two themes was performed across all documents and by issuer type to give an overview of the comparative presence of the themes in the corpus and to assess if sectoral differences or similarities could be discerned.

## 4. FINDINGS

This section will present the findings of the research. While all codes are included in Table 2, the focus here is on the findings resulting from the Sustainability theme. The number and percentage of sources in which the key terms for the Sustainability theme occur are listed in Table 2 and in Figure 1 the codes are ordered by frequency of occurrence across the corpus. 'Sustainability' occurs in more of the documents than any other term in this theme, appearing in 33% of the documents, with a gap to the next highest occurring, 'Africa' and 'Developing World' at 12% and 11% respectively. The least frequently mentioned terms are 'Third World' (3.5%) and 'Low Resource' (2.4%) and associated codes. An entry was returned for each of the search terms around Sustainability. However, in 55% of the documents there were no occurrences of any of the 8 key terms for this theme.

| Search term | No. Documents | Included codes |
|---|---|---|
| Sustainable | 28/84 or 33% | Sustainable, sustainability, sustainably, sustainable development, sustainable society, ecological sustainability, environmental sustainability, sustained participation, agronomic sustainability, unsustainable agriculture, unsustainable, technology [durable, durabilité, sostenibile, sostenibilità] |
| SDG | 7/84 or 8% | SDG, SDGs, Sustainable Development Goal, Sustainable Development Goals [objectifs de développement durable, des ODD, obiettivi di sviluppo sostenibile, OSS] |
| Global South | 5/84 or 6% | Global South, global justice, global gap, global poverty [sud global, sud del mondo] |
| Low/Middle Income | 6/84 or 7% | Low income country, Low income countries, middle income country, middle income countries, low or middle income country, low and middle income countries, LMIC, LMICs [pays à faible revenu, pays à revenu intermédiaire, paesi a basso reddito, paesi a reddito medio] |
| Developing World | 9/84 or 11% | Developing World, developing countries, developing country, developing nation, developing nations, developing economies, emerging economies [monde en développement, mondo/paesi in via di sviluppo] |
| Low Resource | 2/84 or 2% | Low resource country, low resource countries, resource constrained country, resource constrained countries, under-resourced states, resource-poor populations [faible ressources, ressources limitées, risorsa/e bassa, risorse limitate] |
| Africa | 10/84 or 12% | Africa, sub-Saharan Africa, African, African ethics (Ubuntu), South Africa [Afrique, Afrique sub-saharienne, Africaine, Africain, Afrique du Sud, Africa, Africano, Africana, Africa sub-sahariana, Sudafrica] |
| Third World | 3/84 or 4% | Third World, third world countries, third world nations [Tiers-Monde, Pays du tiers-monde, terzo mondo, paesi del terzo mondo] |
| Gender | 45/84 or 54% | Gender, gendered, gendering, genderless, transgender, gender-based [genre, genere] |
| Sex | 24/84 or 29% | Sex, sexism, sexist, sexual harassment, sexual violence, sex-based, sexualised, sexualized, sex robots, sex industry, sex trafficking [sexe, sesso] |
| Women | 27/84 or 32% | Women, trans-women [femmes, donne] |
| Woman | 12/84 or 14% | Woman, trans-woman [femme, donna] |
| Female | 18/84 or 21% | Female, females, feminine [femme, féminin, femmina, femminile] |
| Equality | 55/84 or 65% | Equality, equity, equitable, equal access, equal rights, inequality, inequalities, inequity, inequities [égalité, inégalité, uguaglianza, uaguale, disuguaglianza] |
| Feminism | 0/84 or 0% | Feminism [féminisme, femminismo] |
| Feminist | 2/84 or 2% | Feminist [féministe, femminista] |
| Feminisation | 2/84 or 2% | Feminisation, feminization, feminise, feminize, feminised, feminized |

**Table 2 : Results and Coding for Sustainability and Gender Themes**

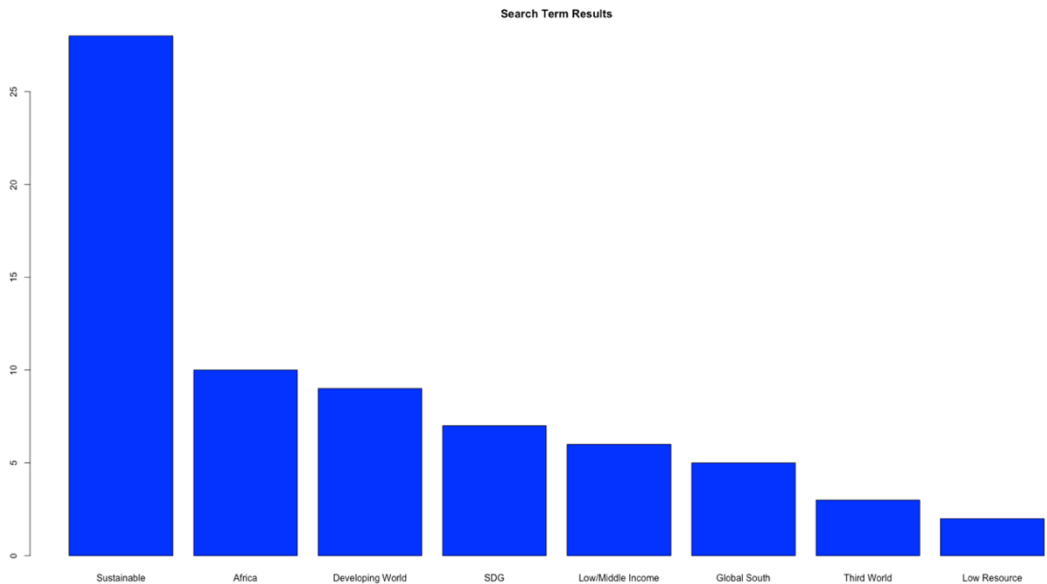

**Figure 1 : Ranked Results for Sustainability Theme**

While 'Sustainable' (and associated codes) occurs most frequently across the documents (in 28 sources), it is also the term with the highest number of mentions (116) in a single document. However, this is an outlier when examined against occurrences of key terms within sources for this theme. Several terms are mentioned rarely within the documents, such as 'Global South' with a maximum of 3 mentions and 'Third World' which never occurs more than once in a source. Analysis of term distribution within sources highlights the low base of occurrences of key terms for this theme. From an examination of the within-document pattern of term occurrences, it is evident that sparse data is a feature, resulting in extremely low median values. A mode of 1 is the most common.

Analysis of results by classification of issuer shows distinct differences between documents from Private sources when compared to those issued by Public and NGO sectors. In Table 3, results for each of the three source categories are shown. Again, while the categorial results for both search themes are included, the focus here is on the Sustainability theme. In the Private documents group, 6 of the 8 search terms associated with sustainability and the Global South are missing. Only 'Sustainable' and 'SDG' occur in any of the 18 documents in this category and these appear in only 3 of the sources, representing 16% of this class. Documents in the Public group do not feature any references to 'Low Resource' and associated codes. In the NGO class, there are occurrences of each of the 8 key terms.

| Search term  | NGO Documents | Private Documents | Public Documents |
|---|---|---|---|
| Sustainable  | 8/35 | 2/18 | 18/31 |
| SDG          | 3/35 | 1/18 | 3/31  |
| Global South | 3/35 | 0/18 | 2/31  |
| Low Income   | 5/35 | 0/18 | 1/31  |
| Developing World | 6/35 | 0/18 | 3/31 |
| Low resource | 2/35 | 0/18 | 0/31 |
| Africa | 6/35 | 0/18 | 4/31 |
| Third World | 2/35 | 0/18 | 1/31 |
| Gender | 23/35 | 5/18 | 17/31 |
| Sex | 14/35 | 0/18 | 10/31 |
| Women | 13/35 | 1/18 | 13/31 |
| Woman | 6/35 | 1/18 | 5/31 |
| Female | 9/35 | 0/18 | 9/31 |
| Equality | 27/35 | 6/18 | 22/31 |
| Feminism | 0/35 | 0/18 | 0/31 |
| Feminist | 2/35 | 0/18 | 0/31 |
| Feminisation | 1/35 | 0/18 | 1/31 |

**Table 3 : Theme Results per Category**

Results from the terms present in the documents differ across issuer types, especially between those from the Private and the other two issuing sectors. Found most frequently in each category of documents, 'Sustainability' occurs in 58% of Public documents. This is much higher than in the 23% of the NGO sources, which again is higher than the 11% total for the Private category. The second most frequently referenced key term for both NGO and Public classes is 'Africa', appearing in 17% and 13% of documents in their respective categories. Notably, these second ranked terms are at a higher percentage than the first ranked term in the Private class. Within the Private grouping, the second and only other occurring key term 'SDG' is found in 5.5% of documents.

Regarding the occurrence of individual terms, there are some similarities between the NGO and Public groups, but they are not as obvious as in the search relating to gender and feminism. The distribution of search terms across documents shows some parallels between sources issued by the NGO and Public sectors. However, the distribution within the Private grouping of documents is very different to the other two categories. When corrected for scale, the gap between NGO and Public documents on one hand and Private on the other, becomes clear. While Public sources scale from 0% to almost 60%, NGO documents have a narrower range from 6% to 23%, while the values for the Private category are restricted to a scale spanning 5.5% to 11%. This will be discussed in the analysis in the next section.

## 5. DISCUSSION

The results presented in the previous section reveal a dearth of references to both gender and sustainability in the AI ethics documents, especially those issued by the private sector, which could be indicative of a concerning imbalance in the global discourse on ethical AI. If the corpus analysed here is representative of the broader landscape, then these findings could be considered evidence of a worrying absence of significant voices in the AI ethics debate, specifically those often most marginalised by the technology, namely women and the Global South. If the debate is being shaped disproportionately by higher-income countries and a largely male-dominated industry, are gender diversity, global fairness and cultural pluralism being neglected?

Perhaps the lack of consideration of sustainability and gender found in this study is rooted in the geographical origin of the documents analysed. From Figure 2[1], the dominance of ethics sources from more economically developed countries and a glaring absence of documents from the Global South is evident. Of a corpus of 84, the US contributes 21 sources (25%) and 13 (15%) originate in the United Kingdom (UK): these two countries therefore provide 40% of the documents analysed. This could be explained by a language bias towards English in the Jobin et al. (2019) study. However, it would appear from other synthesis papers and surveys that the proliferation of AI ethics grey literature is coming from the Global North/ Western world (Fjeld et al., 2020; Floridi et al., 2018; Morley et al., 2020).

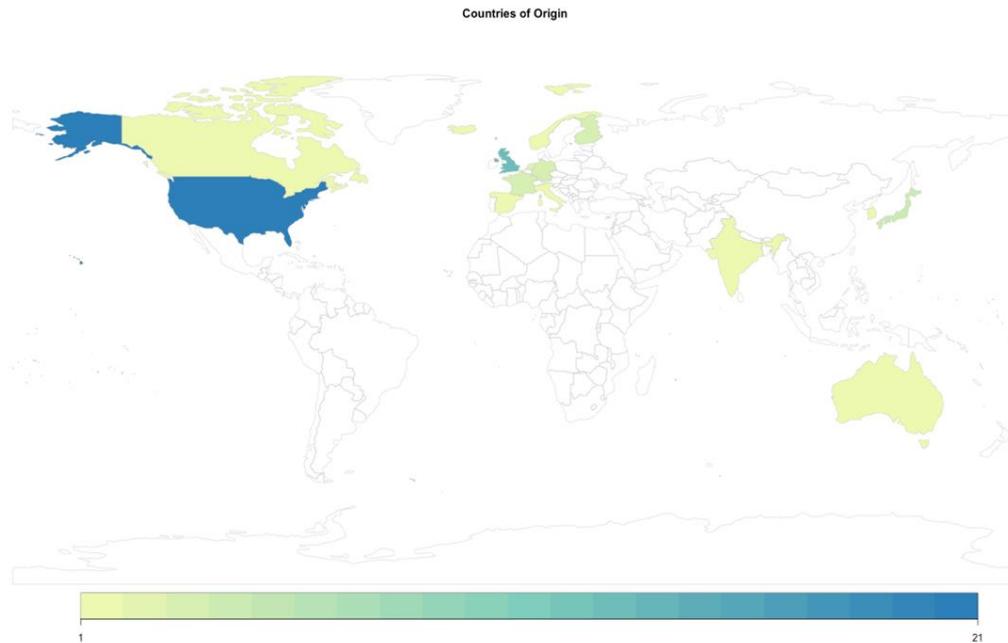

**Figure 2 : AI Ethics Documents in Corpus by Country (N=84)**

Furthermore, this phenomenon is again visible following an examination of the Council of Europe's (CoE) data visualisation of AI initiatives[2]. This site, which collates documents relating to AI had 456 documents at the date of access (March 2021), sourced from think tanks, professional associations, civil society, academia, international agencies, national governments or authorities, multi-stakeholder programmes and the private sector. While not exhaustive, this collection is nonetheless telling, in that of the over 450 sources, only 3 are from the Global South. It would appear therefore that not all global regions are participating equally in the AI ethics discourse. Notably, the most common concepts found in the CoE documents relate to privacy, human rights, transparency, responsibility, trust and accountability. Although sustainability and the SDGs are found as concepts in the sources, they appear with much less frequency and there is no reference to gender as a concept within the documents.

Given that the documents in the Jobin et al. (2019) corpus used for this survey are concerned with a variety of topics within the broad sphere of AI, it is perhaps not surprising to find a lack of coverage of the gender and sustainability themes. Although the sources focus on such diverse topics as robotics, radiology, automated driving, data analytics and AI for business, all of these documents are part of a corpus of literature primarily concerned with the ethics of AI. The absence of any real consideration of these themes across the grey literature is a troubling gap. With an ever-growing body of literature on AI ethics, which has a potentially normative influence, it is concerning that certain topics are not only lacking prominence in the discourse but are paid so little attention that they are not deemed worthy of ethical concern.

---

[1] The map does not include documents classified as 'International' or 'N/A'.
[2] https://www.coe.int/en/web/artificial-intelligence/national-initiatives

What are the social and ethical implications of not having a more inclusive AI ethics landscape? A lack of attention in ethical debates around AI, to concerns pertinent to sustainability and the Global South, is problematic. If these are not part of the mainstream discussion, they are unlikely to be featured in public debate or policy development. As a consequence, considerations around such issues as the environmental impact associated with AI computational resources requiring vast energy sources might be overlooked and decisions made which can have a devastating effect on communities who are already vulnerable such as those in sub-Saharan Africa. Moreover, the sectoral results in this study indicate that the Private sector, in particular, has a lacuna when it comes to sustainability concerns which has crucial implications for those in the Global South where multinational corporations are prominent in infrastructural investment.

Equally, if the voices of women and the Global South are not included in the grey literature, they are likely to be ignored by policy makers as well as multinational organisations. If such issues are not part of the current discourse around AI ethics, there is a great risk of AI impeding gender equality and reinforcing paternalistic ideologies. With an emerging focus on fairness in AI, it is, as Leavy observes, "*essential that women are at the core of who defines the concept of fairness*" (Leavy, 2018). However, if women are not at the table and not considered by those setting the agenda, progress already achieved in gender equality, sustained by feminist thought, could be undermined. Furthermore, narrow representation at the level of grey literature is reflective of not only a geographical dominance but also of an imbalance of certain social and demographic groups. AI and the ethical framework that supports it could better reflect the diversity of the global community by addressing the power imbalance within international discourse on the ethics of AI.

## 6. CONCLUSIONS

This study assessed a body of AI ethics documents collected by Jobin et al. (2019) and categorised by issuer-type (NGO, Private, Public), for inclusion of 17 key terms associated with sustainability and gender. Findings reveal a dearth of references to these themes in the AI ethics documents, especially those issued by the Private sector. While these documents reflect a breadth of ethical and social issues around AI, it would appear that gender and the Global South are neglected. These results, as interpreted here, indicate a lack of attention to inclusivity in the framing of the ethical discourse around AI. The sparsity of data across the literature, regardless of sector, is suggestive of a potentially concerning imbalance in the global discourse on AI ethics where women and the Global South are underrepresented. This presents a double disadvantage: while gender inequality is a global problem, it can be particularly so in the Global South.

While there is still much to learn, these findings help to reveal a lack of inclusivity in the ethical framing of key issues in AI. Such findings have implications for policymakers, international organisations, technology companies and all working to design, develop and implement AI in a fair and ethical manner. Any exclusion of significant voices means much of the globe (and half the world's population) could be forced to use and adapt to technologies developed without their input but also guided by an ethical framework which may not reflect their real, lived experience or the reality of the social, cultural, political, environmental, and ethical environments in which they exist. Continued underrepresentation has implications for the reinforcement of sexist, paternalistic, ethnocentric and colonial ideologies and associated power structures.

Although it is disappointing that these AI ethical and governance documents do not take appropriate account of the importance of gender or sustainability, it is still possible to inform discourse and shape the future of emerging AI policy. Identifying and understanding the power structures in play in AI ethics is an important area for research. Future work is required around how best to challenge existing systems and approaches to bring better balance and improved inclusivity to the AI ethical discourse. This could be through applications of feminist, intersectional, decolonial and post-colonial thought. Bringing these differing lenses to the AI ethical debate could help influence the direction of AI ethics in a more inclusive and participatory way.

# ACKNOWLEDGEMENTS

This research was conducted at the ADAPT SFI Research Centre at Trinity College Dublin. The ADAPT SFI Centre for Digital Content Technology is funded by Science Foundation Ireland through the SFI Research Centres Programme and is co-funded under the European Regional Development Fund (ERDF) through Grant #13/RC/2106_P2.

# APPENDIX A

### Ethics Principles by Country of Issuer and Category

| Document/website | Issuing Agency/Org | Country | Category |
|---|---|---|---|
| Artificial Intelligence. Australia's Ethics Framework: A Discussion Paper | Department of Industry Innovation and Science | Australia | Public |
| Montréal Declaration: Responsible AI | Université de Montréal | Canada | Public |
| AI4People—An Ethical Framework for a Good AI Society: Opportunities, Risks, Principles, and Recommendations | AI4People | EU | NGO |
| Position on Robotics and Artificial Intelligence | The Greens (Green Working Group Robots) | EU | Public |
| Report with Recommendations to the Commission on Civil Law Rules on Robotics | European Parliament | EU | Public |
| Ethics Guidelines for Trustworthy AI | High-Level Expert Group on Artificial Intelligence | EU | Public |
| European Ethical Charter on the Use of Artificial Intelligence in Judicial Systems and Their Environment | Council of Europe: European Commission for the Efficiency of Justice (CEPEJ) | EU | Public |
| Statement on Artificial Intelligence, Robotics and 'Autonomous' Systems | European Commission, European Group on Ethics in Science and New Technologies | EU | Public |
| Tieto's AI Ethics Guidelines | Tieto | Finland | Private |
| Commitments and Principles https://www.op.fi/op-financial-group/corporate-social-responsibility/commitments-and-principles | OP Group | Finland | Private |
| Work in the Age of Artificial Intelligence. Four Perspectives on the Economy, Employment, Skills and Ethics | Ministry of Economic Affairs and Employment | Finland | Public |
| How Can Humans Keep the Upper Hand? Report on the Ethical Matters Raised by AI Algorithms | French Data Protection Authority (CNIL) | France | Public |
| For a Meaningful Artificial Intelligence. Towards a French and European Strategy | Mission Villani | France | Public |
| Ethique de la Recherche en Robotique | CERNA (Allistene) | France | Public |
| AI Guidelines | Deutsche Telekom | Germany | Private |
| SAP's Guiding Principles for Artificial Intelligence | SAP | Germany | Private |

| Automated and Connected Driving: Report | Federal Ministry of Transport and Digital Infrastructure, Ethics Commission | Germany | Public |
|---|---|---|---|
| Ethics Policy https://www.iiim.is/ethics-policy/ | Icelandic Institute for Intelligent Machines (IIIM) | Iceland | NGO |
| Discussion Paper: National Strategy for Artificial Intelligence | National Institution for Transforming India (NITI Aayog) | India | Public |
| Artificial Intelligence and Machine Learning: Policy Paper | Internet Society | International | NGO |
| Ethical Principles for Artificial Intelligence and Data Analytics | Software & Information Industry Association (SIIA), Public Policy Division | International | NGO |
| ITI AI Policy Principles | Information Technology Industry Council (ITI) | International | NGO |
| Ethically Aligned Design. A Vision for Prioritizing Human Well-being with Autonomous and Intelligent Systems, Version 2 | Institute of Electrical and Electronics Engineers (IEEE), The IEEE Global Initiative on Ethics of Autonomous and Intelligent Systems | International | NGO |
| Top 10 Principles for Ethical Artificial Intelligence | UNI Global Union | International | NGO |
| The Malicious Use of Artificial Intelligence: Forecasting, Prevention, and Mitigation | Future of Humanity Institute; University of Oxford; Centre for the Study of Existential Risk; University of Cambridge; Center for a New American Security; Electronic Frontier Foundation; OpenAI | International | NGO |
| White Paper: How to Prevent Discriminatory Outcomes in Machine Learning | WEF, Global Future Council on Human Rights 2016-2018 | International | NGO |
| The Toronto Declaration: Protecting the Right to Equality and Non-discrimination in Machine Learning Systems | Access Now; Amnesty International | International | NGO |
| Report of COMEST on Robotics Ethics | COMEST/UNESCO | International | Public |
| Privacy and Freedom of Expression in the Age of Artificial Intelligence | Privacy International & Article 19 | International | NGO |
| Artificial Intelligence: Open Questions About Gender Inclusion | W20 | International | NGO |
| Universal Guidelines for Artificial Intelligence | The Public Voice | International | NGO |
| Ethics of AI in Radiology: European and North American Multisociety Statement | American College of Radiology; European Society of Radiology; Radiology Society of North America; Society for Imaging Informatics in Medicine; European Society of Medical | International | NGO |
| | Imaging Informatics; Canadian Association of Radiologists; American Association of Physicists in Medicine | | |
| Ethically Aligned Design: A Vision for Prioritizing Human Well-being with Autonomous and Intelligent Systems, First Edition (EAD1e) | Institute of Electrical and Electronics Engineers (IEEE), The IEEE Global Initiative on Ethics of Autonomous and Intelligent Systems | International | NGO |
| Charlevoix Common Vision for the Future of Artificial Intelligence | Leaders of the G7 | International | Public |
| Declaration on Ethics and Data Protection in Artificial Intelligence | ICDPPC | International | Public |
| L'intelligenza Artificiale al Servizio del Cittadino | Agenzia per l'Italia Digitale (AGID) | Italy | Public |
| Sony Group AI Ethics Guidelines | Sony | Japan | Private |
| The Japanese Society for Artificial Intelligence Ethical Guidelines | Japanese Society for Artificial Intelligence Japan | Japan | Public |
| Report on Artificial Intelligence and Human Society (unofficial translation) | Advisory Board on Artificial Intelligence and Human Society (initiative of the Minister of State for Science and Technology Policy) | Japan | Public |
| Draft AI R&D Guidelines for International Discussions | Institute for Information and Communications Policy (IICP), The Conference toward AI Network Society | Japan | Public |
| Tenets | Partnership on AI | N/A | NGO |
| Principles for Accountable Algorithms and a Social Impact Statement for Algorithms | Fairness, Accountability, and Transparency in Machine Learning (FATML) | N/A | NGO |
| 10 Principles of Responsible AI | Women Leading in AI | N/A | NGO |
| Human Rights in the Robot Age Report | The Rathenau Institute | Netherlands | NGO |
| Dutch Artificial Intelligence Manifesto | Special Interest Group on Artificial Intelligence (SIGAI), ICT Platform Netherlands (IPN) | Netherlands | NGO |
| Artificial Intelligence and Privacy | The Norwegian Data Protection Authority | Norway | Public |
| Discussion Paper on Artificial Intelligence (AI) and Personal Data —Fostering Responsible Development and Adoption of AI | Personal Data Protection Commission Singapore | Singapore | Public |
| Mid- to Long-Term Master Plan in Preparation for the Intelligent Information Society | Government of the Republic of Korea | South Korea | Public |
| AI Principles of Telefónica | Telefónica | Spain | Private |
| AI Principles & Ethics | Smart Dubai | UAE | Public |
| DeepMind Ethics & Society Principles (not avail at link in Jobin paper, instead looked at | DeepMind Ethics & Society | UK | NGO |
| themes) https://deepmind.com/about/ethics-and-society#themes | | | |
| Business Ethics and Artificial Intelligence | Institute of Business Ethics | UK | NGO |
| Ethics Framework: Responsible AI | Machine Intelligence Garage Ethics Committee | UK | NGO |
| Machine Learning: The Power and Promise of Computers that Learn by Example | The Royal Society | UK | NGO |
| Ethical, Social, and Political Challenges of Artificial Intelligence in Health | Future Advocacy | UK | NGO |
| The Ethics of Code: Developing AI for Business with Five Core Principles | Sage | UK | Private |
| The Responsible AI Framework (used updated version A Practical Guide to Responsible AI) | PricewaterhouseCoopers | UK | Private |
| Responsible AI and Robotics. An Ethical Framework. (used Responsible AI: A Framework for Building Trust in your AI Solutions) | Accenture UK | UK | Private |
| Principles of robotics https://epsrc.ukri.org/research/ourportfolio/themes/engineering/activities/principlesofrobotics/ | Engineering and Physical Sciences Research Council UK (EPSRC) | UK | Public |
| Big Data, Artificial Intelligence, Machine Learning and Data Protection | Information Commissioner's Office | UK | Public |
| AI in the UK: Ready, Willing and Able? | UK House of Lords, Select Committee on Artificial Intelligence | UK | Public |
| Artificial Intelligence (AI) in Health | Royal College of Physicians | UK | Public |
| Initial Code of Conduct for Data-Driven Health and Care Technology (used updated version 19 Jan 2021) | UK Department of Health & Social Care | UK | Public |
| Unified Ethical Frame for Big Data Analysis. IAF Big Data Ethics Initiative, Part A | The Information Accountability Foundation | USA | NGO |
| The AI Now Report. The Social and Economic Implications of Artificial Intelligence Technologies in the Near-Term (2016) | AI Now Institute | USA | NGO |
| Statement on Algorithmic Transparency and Accountability (2017) | Association for Computing Machinery (ACM) | USA | NGO |
| AI Principles https://futureoflife.org/ai-principles/ | Future of Life Institute | USA | NGO |
| Policy Recommendations on Augmented Intelligence in Health Care H-480.940 | American Medical Association (AMA) | USA | NGO |
| Governing Artificial Intelligence. Upholding Human Rights & Dignity | Data & Society | USA | NGO |
| Digital Decisions | Center for Democracy & Technology | USA | NGO |
| Science, Law and Society (SLS) Initiative https://thefuturesociety.org/2017/07/15/principles-law-and-society-initiative/ | The Future Society | USA | NGO |
| AI Now 2018 Report | AI Now Institute | USA | NGO |

| AI Now 2017 Report | AI Now Institute | USA | NGO |
|---|---|---|---|
| AI—Our Approach (not at link in Jobin's paper, instead used below) https://www.microsoft.com/en-us/ai/responsible-ai?activetab=pivot1%3aprimaryr6 | Microsoft | USA | Private |
| Artificial Intelligence. The Public Policy Opportunity (2017) | Intel Corporation | USA | Private |
| IBM's Principles for Trust and Transparency | IBM | USA | Private |
| Open AI Charter | Open AI | USA | Private |
| Everyday Ethics for Artificial Intelligence. A Practical Guide for Designers and Developers | IBM | USA | Private |
| Intel's AI Privacy Policy Paper. Protecting Individuals' Privacy and Data in the Artificial Intelligence World | Intel Corporation | USA | Private |
| Introducing Unity's Guiding Principles for Ethical AI—Unity Blog | Unity Technologies | USA | Private |
| Responsible Bots: 10 Guidelines for Developers of Conversational AI | Microsoft | USA | Private |
| Preparing for the Future of Artificial Intelligence | Executive Office of the President; National Science and Technology Council; Committee on Technology | USA | Public |
| The National Artificial Intelligence Research and Development Strategic Plan | National Science and Technology Council; Networking and Information Technology Research and Development Subcommittee | USA | Public |